# Spin- to charge-current conversion in altermagnetic candidate $RuO_2$ probed by terahertz emission spectroscopy

J. Jechumtál[1], O. Gueckstock[2], K. Jasenský[1], Z. Kašpar[1,3], K Olejník[3], M. Gaerner[4], G. Reiss[4], S. Moser[5], P. Kessler[5], G. De Luca[6], S. Ganguly[7], J. Santiso[7], D. Scheffler[3], J. Zázvorka[1], P. Kubaščík[1], H. Reichlova[3], E. Schmoranzerova[1], P. Němec[1], T. Jungwirth[3,8], P. Kužel[3], T. Kampfrath[2] and L. Nádvorník[1*]

1. Faculty of Mathematics and Physics, Charles University, 121 16 Prague, Czech Republic
2. Department of Physics, Freie Universität Berlin, 14195 Berlin, Germany
3. Institute of Physics, Czech Academy of Science, 16200 Prague, Czech Republic
4. Bielefeld University, Faculty of Physics, 33615 Bielefeld, Germany
5. Physikalisches Institut and Würzburg-Dresden Cluster of Excellence ct.qmat, Universität Würzburg, 97074 Würzburg, Germany
6. Materials Science Institute of Barcelona (ICMAB-CSIC), 08193 Barcelona, Spain
7. Catalan Institute of Nanoscience and Nanotechnology, 08193 Barcelona, Spain
8. School of Physics and Astronomy, University of Nottingham, UK

* E-mail: lukas.nadvornik@matfyz.cuni.cz

## Abstract

Using the THz emission spectroscopy, we investigate ultrafast spin-to-charge current conversion in epitaxial thin films of the altermagnetic candidate $RuO_2$. We perform a quantitative analysis of competing effects that can contribute to the measured anisotropic THz emission. These include the anisotropic inverse spin splitter and spin Hall effects in $RuO_2$, the anisotropic conductivity of $RuO_2$, and the birefringence of the $TiO_2$ substrate. We observe that the leading contribution to the measured signals comes from the anisotropic inverse spin Hall effect, with an average spin-Hall angle of $2.4 \cdot 10^{-3}$ at room temperature. In comparison, a possible contribution from the altermagnetic inverse spin-splitter effect is found to be approximately $2-4 \cdot 10^{-4}$. Our work stresses the importance of carefully disentangling spin-dependent phenomena that can be generated by the unconventional altermagnetic order, from the effects of the relativistic spin-orbit coupling.

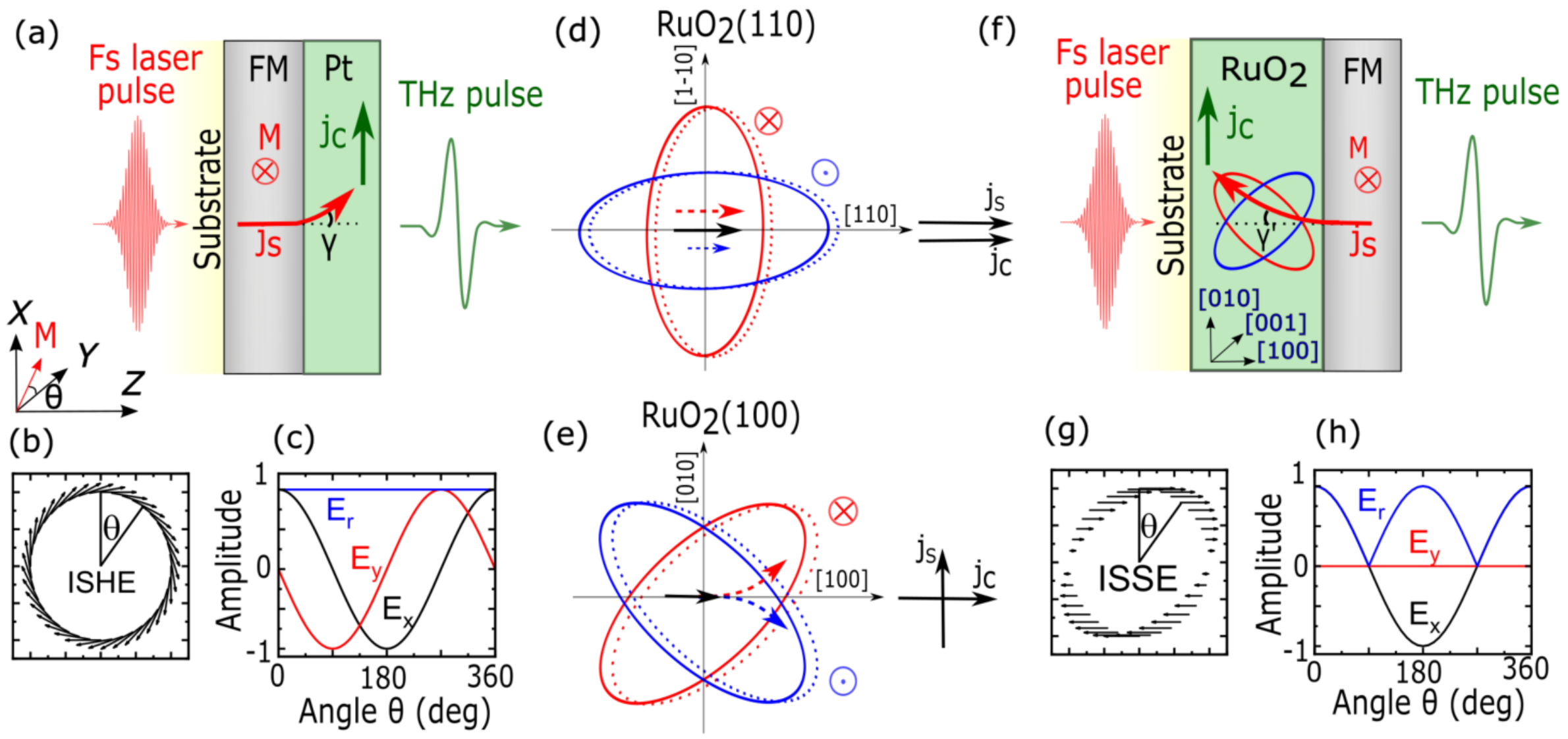

FIG. 1. THz emission from FM|Pt, FM|$RuO_2$ and symmetry of spin splitter effect (SSE). (a) Spintronic terahertz emission from a reference bilayer ferromagnet (FM)|Pt. A femtosecond laser pulse excites the FM with in-plane magnetization $\boldsymbol{M}$ and, thus, injects a spin-current $\boldsymbol{j_s}$ (red arrow) into Pt, where it is converted into a transverse charge current $\boldsymbol{j_c}$ (green arrow) by the ISHE, with an efficiency defined by the angle $\boldsymbol{\gamma}$. Consequently, a THz electromagnetic pulse is radiated. When changing the angle $\boldsymbol{\theta}$ of $\boldsymbol{M}$ in the x-y plane by an external magnetic field, (b) the emitted electric field $\boldsymbol{E}$ is always perpendicular to $\boldsymbol{M}$ and has identical amplitudes and waveforms, resulting in (c) the typical harmonic projections $\boldsymbol{E_x}$, $\boldsymbol{E_y}$, and the constant total magnitude $\boldsymbol{E_r}$. (d), (e) Symmetry of altermagnetic SSE. (d) Sending $\boldsymbol{j_c}$ along [110] crystal direction of a d-wave altermagnet leads to spin polarization of current, but no generation of any transversal component of $\boldsymbol{j_s}$ to the direction of $\boldsymbol{j_c}$. (e) $\boldsymbol{j_c}$ sent along [100] crystal direction leads to spin-dependent deflection of up and down spins, which translates to the transversal $\boldsymbol{j_s}$. The inverse equivalent, where $\boldsymbol{j_s}$ produces transversal $\boldsymbol{j_c}$, is dubbed inverse spin splitter effect (ISSE). (f) THz emission from $RuO_2$(100)|FM with potential altermagnetic ISSE, acting analogously to ISHE in (a). $\boldsymbol{\theta} = \boldsymbol{0}$ corresponds to $\boldsymbol{M} \parallel [\boldsymbol{001}]$-direction of $RuO_2$. (g,h) Analogous projections of the ISSE-driven emitted $\boldsymbol{E}$. Unlike (b,c), $\boldsymbol{E_y} = \boldsymbol{0}$ because $\boldsymbol{y}$-axis is set along the [001]-direction of $RuO_2$. This is also the direction in which the Neél vector lies and, consequently, no spin splitting is expected. Moreover, $\boldsymbol{E_r}$ is not constant as the projection of the injected spin to [001]-direction of $RuO_2$ is varying with angle $\boldsymbol{\theta}$.

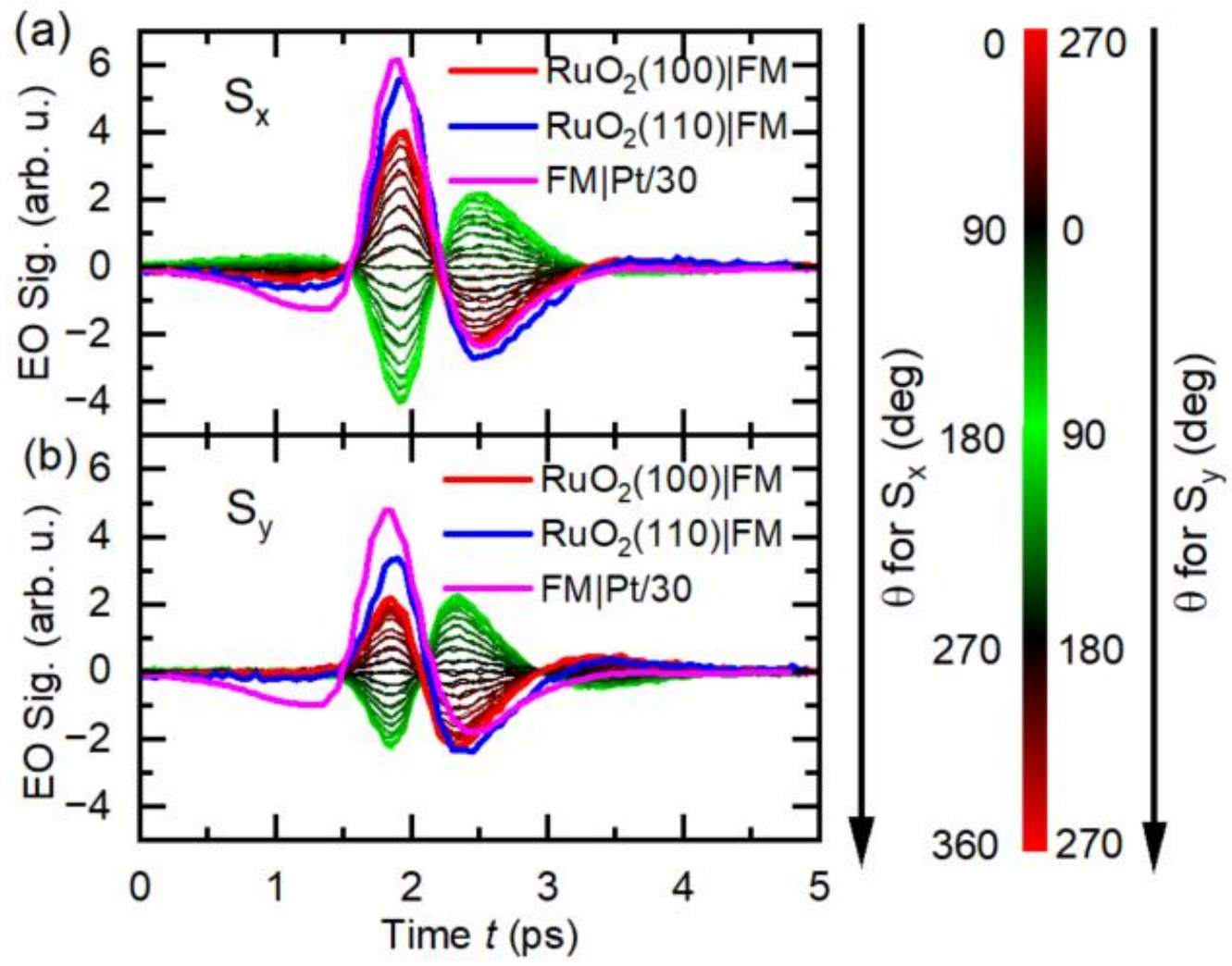

FIG 2. Typical THz-emission signals. (a) The $\boldsymbol{x}$-projection $\boldsymbol{S_x(t)}$ of the EO signals from $RuO_2$(100)|CoFeB (red), $RuO_2$(110)|CoFeB (blue), and reference sample CoFeB|Pt (magenta, rescaled by an indicated factor of 30) for $\boldsymbol{\theta = 0}$°. (b) The same dataset for $\boldsymbol{S_y(t)}$ and $\boldsymbol{\theta = 270}$°. An example of the typical full $\boldsymbol{\theta}$-dependence is shown for (100)-$RuO_2$|CoFeB (sequence of 40 colors shown in color map on right side).

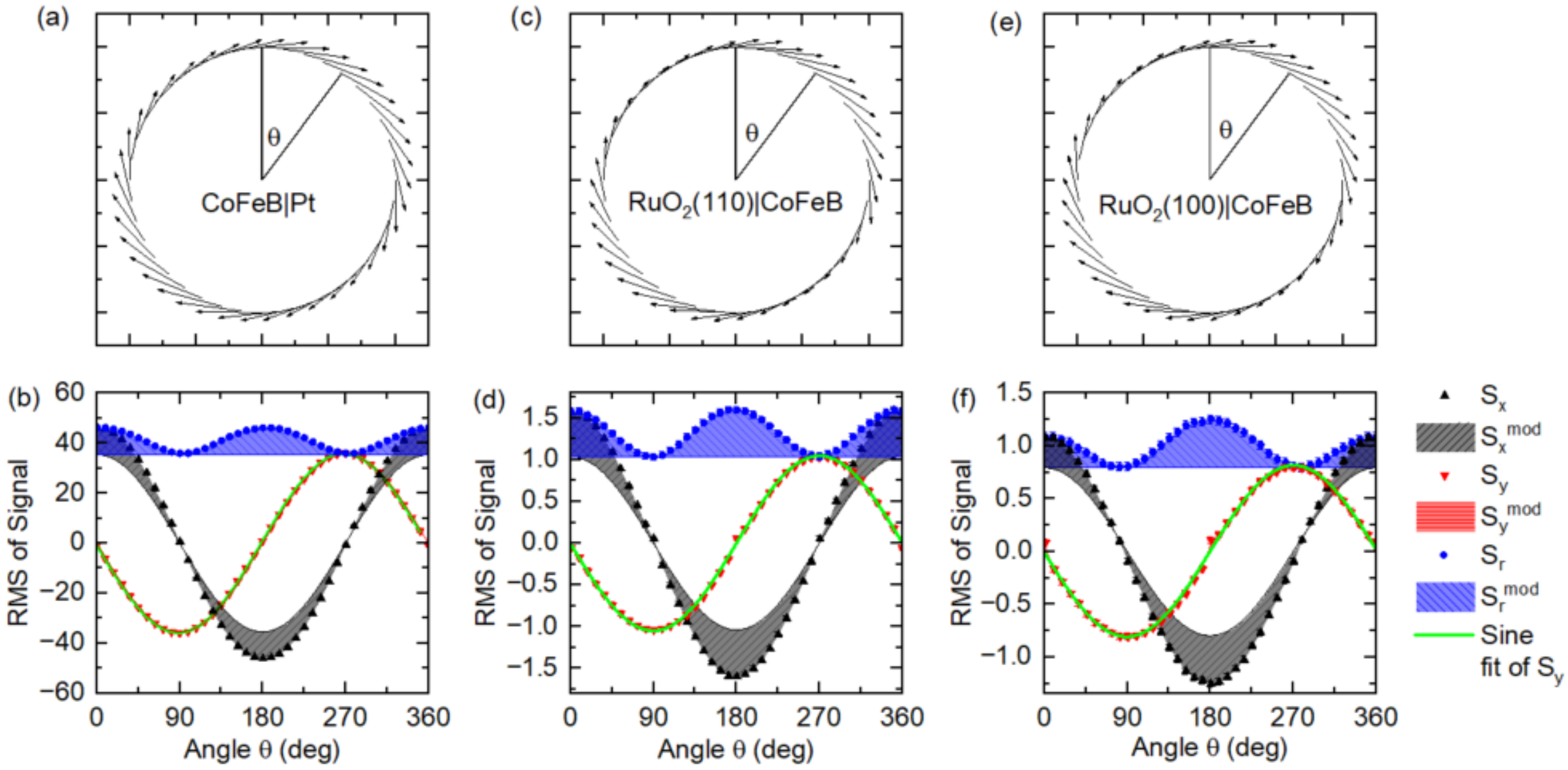

FIG. 3. Analysis of $\boldsymbol{\theta}$-dependence. (a) RMS of $\boldsymbol{S_x}$ and $\boldsymbol{S_y}$ as a function of $\boldsymbol{\theta}$ as a texture plot for the CoFeB|Pt reference sample. (b) RMS of the measured components $\boldsymbol{S_x}(\boldsymbol{\theta})$, $\boldsymbol{S_y}(\boldsymbol{\theta})$ and of the magnitude $\boldsymbol{S_r}(\boldsymbol{\theta})$ (symbols) compared with the model (lines) by Eq. (1) using only the isotropic ISHE contribution. Here, $\boldsymbol{S_y}(\boldsymbol{\theta})$ is fit by $\boldsymbol{A \sin \theta}$ [Eq. (1), green curve], and the shaded areas for $\boldsymbol{S_x}(\boldsymbol{\theta})$ and $\boldsymbol{S_r}(\boldsymbol{\theta})$ highlight the difference between the experiment and the isotropic model. (c), (d) Analogous analysis for $RuO_2$(110)|CoFeB, where no ISSE is expected, and (e), (f) for $RuO_2$(100)|CoFeB sample, where the ISSE is allowed. Notice that textures in panels (a), (c) and (e) are based on the same underlying data as shown in panels (b), (d) and (f), both providing different views on signal symmetries and their amplitudes.

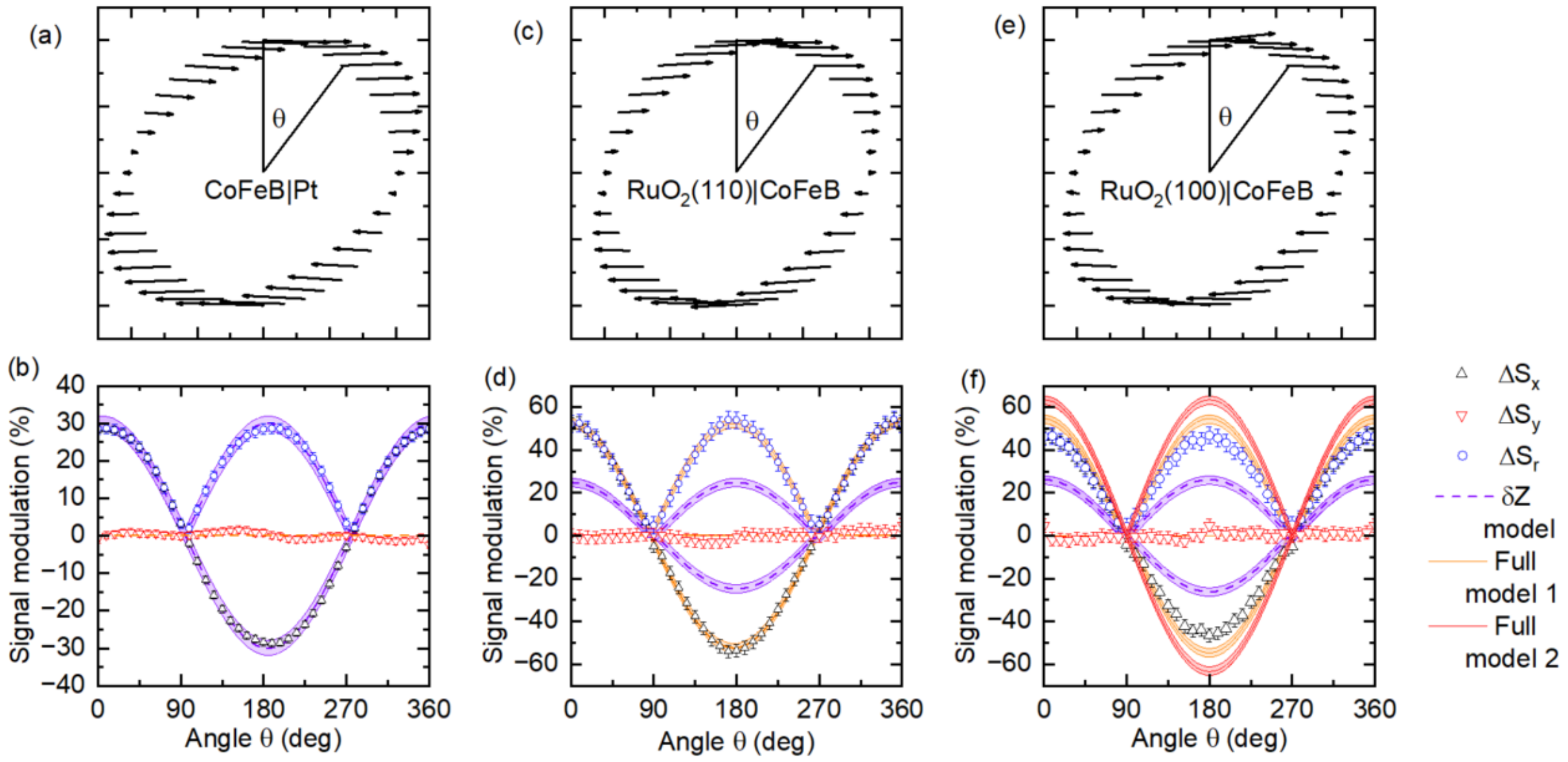

FIG. 4. Modulation of $\boldsymbol{\theta}$-dependence. Panels (a) and (b): CoFeB|Pt sample. (a) Emission texture after subtraction of isotropic ISHE contribution. (b) Residuals $\boldsymbol{\Delta S_x(\theta)}$, $\boldsymbol{\Delta S_y(\theta)}$, $\boldsymbol{\Delta S_r(\theta)}$ from (a), normalized to $\boldsymbol{A}$ (symbols), compared to the model of anisotropic outcoupling part of Eq. (2), i.e, only $\boldsymbol{\delta Z/Z_y \neq 0}$ (violet dashed curve). Panels (c), (d): Analogous analysis for (110)-$RuO_2$|CoFeB sample, the ISSE inactive orientation. Panels (e), (f): Analogous for (100)-$RuO_2$|CoFeB sample, the ISSE active orientation. In (d) and (f), we complement the partial model (only $\boldsymbol{\delta Z/Z_y \neq 0}$, violet dashed curve) with the full models by whole Eq. (2) including anisotropic ISHE. Model 1 assumes different spin transparency between the two $RuO_2$ samples (orange curve), Model 2 assumes different conversion in the FM layer (red curve). See the main text for detailed description of the modeling procedure. Notice that textures in panels (a), (c) and (e) are based on the same underlying data as shown in panels (b), (d) and (f).

Altermagnetism has recently emerged as a new symmetry class in magnetism [1], characterized by a momentum-dependent alternating spin polarization in the reciprocal space of collinear, magnetically compensated systems. Unlike conventional ferromagnets or antiferromagnets, the altermagnetic order spontaneously breaks both spin-space and real-space rotation symmetries, while preserving a symmetry combining spin-space and real-space rotation transformations. This property gives rise to unconventional spin transport, such as the spin-splitter effect, giant magneto-transport anisotropies, and spin-polarized currents without net magnetization [2,3]. Among the broad class of proposed altermagnets, rutile-structured $RuO_2$ stands out as a particularly compelling candidate [2,4,5]. In the altermagnetic phase, it is predicted to combine metallic conductivity with *d*-wave-type spin order, all in a simple binary compound [1-5], making it an ideal platform for both theoretical and experimental investigations.

$RuO_2$ has long been known as a non-magnetic, versatile material, useful in electrochemistry and electronics [6], yet it has been going through a dynamic experimental development in recent years. After reports indicating a presence of a compensated collinear magnetic order, based on neutron diffraction [7] and X-ray scattering [8], several transport experiments showed signals suggestive of a theoretically predicted [4] altermagnetic anomalous Hall [9], a theoretically predicted [10] spin-splitter effect (SSE) measured via a GHz ferromagnetic resonance [11,12,13], and an inverse spin-splitter effect (ISSE) detected by a DC spin Seebeck effect [14] of by THz emission [15]. However, the consensus on the magnetic ordering of $RuO_2$ has been recently challenged by muon spin rotation [16], neutron scattering [17] and angle-resolved photoemission spectroscopy (ARPES) [18] studies reporting no magnetic order in the studied samples. The ongoing debate on altermagnetism in $RuO_2$ involves, on one hand, additional observations attributed to the altermagnetic order, including the tunneling magnetoresistance [19], the anomalous Hall effect [20,21], the spin-split band structure measured by ARPES [22], the DC SSE-induced switching of an adjacent magnet [23], or the optically-induced spin-polarization [24]. On the other hand, recent thermal-transport [25] and spin-pumping [26] studies attributed the observed spin-to-charge conversion to an anisotropic inverse spin Hall effect (ISHE), rather than the ISSE. Similarly, a recent THz-emission study did not find evidence for the ISSE, and ascribed the observed signals to the anisotropic conductivity of $RuO_2$ [27]. Therefore, it is crucial for the current discussion to scrutinize experimental data by considering all other competing effects that may mimic the altermagnetic (I)SSE by sharing analogous experimentally observable symmetries.

In this letter, we provide an ultrafast probe of the spin-to-charge current conversion in $RuO_2$-based multilayers by the time-domain THz emission spectroscopy. It is a natural tool for probing processes with expected ultrafast dynamics, as it can infer, unlike transport and low- frequency techniques, the dynamics and potential speed limits of studied phenomenon [28,29,30]. By considering three possible sources of competing signals to the anisotropic ISSE – the anisotropic ISHE in $RuO_2$, the anisotropic conductivity of $RuO_2$, and the birefringence of the $TiO_2$ substrate, we analyze the strongly anisotropic THz emission in a quantitative way by a comparison to the emission model. The nearly perfect match of our calculations with the measured data allows us to infer the magnitude and polarity of the isotropic part of the ISHE and to demonstrate a strong, anisotropic component of the ISHE in the THz range. From our analysis of the measured data at room temperature, we conclude that the ISSE angle is approximately $2-4\cdot 10^{-4}$. This is three orders of magnitude smaller than the zero-temperature theoretical value considering a single domain altermagnetic state of $RuO_2$ [10,2].

The experiment follows the emission scheme from spintronic THz emitters [Fig. 1(a)] [31]. Here, an ultrashort laser pulse excites a bilayer consisting of a ferromagnetic-metal layer FM and a heavy-metal layer HM, for example Pt. The optical heating generates a transient spin voltage in FM and, thus, drives a spin current with density $\boldsymbol{j}_s$ across the FM/HM interface [32]. The spin orientation of $\boldsymbol{j}_s$ is determined by the direction of the in-plane magnetization $\boldsymbol{M}$ of FM. Inside HM, $\boldsymbol{j}_s$ is converted to an in-plane charge current with density $\boldsymbol{j}_c$ via the ISHE. Finally, the total emitted field $\boldsymbol{E}$ directly behind the sample corresponds to the total integrated charge current $\boldsymbol{I}_c = \int \boldsymbol{j}_c(z)dz$ [30,33]:

$$\boldsymbol{E} = eZ\boldsymbol{I}_c \propto eZ\gamma\boldsymbol{I}_s \times \boldsymbol{e}_{\boldsymbol{M}}, \quad (1)$$

where $e$ is the elementary charge, $Z$ is the sample impedance, $\gamma$ is the spin-Hall angle, $\boldsymbol{I_s} = \int \boldsymbol{j_s}(z)dz$ is the total integrated spin current, $\boldsymbol{e_M}$ is the unit vector in the direction of $\boldsymbol{M}$, and the ISHE is isotropic. It follows from the cross product in Eq. (1) that the emitted $\boldsymbol{E}$ is perpendicular to $\boldsymbol{M}$ and forms an isotropic texture [Fig. 1(b)] if $\boldsymbol{M}$ is rotated by an angle $\theta$. Correspondingly, the x- and y- components of $\boldsymbol{E}$ follow the harmonic $\theta$-dependence, $E_x \propto \cos\theta$ and $E_y \propto -\sin\theta$, while the magnitude $E_r = |\boldsymbol{E}|$ is constant [Fig. 1(c)].

If we replace HM = Pt by a suitably oriented d-wave altermagnet, in our case (100)-grown $RuO_2$ in the altermagnetic phase, we expect the ISSE [Fig. 1(d) and (e)]. This phenomenon, together with its complementary spin-polarization effect, are theoretically predicted to arise from a spin-split band-structure the polarity of which alternates in the [110] and [1-10] crystal directions of $RuO_2$ [10]. The spin-polarizer effect arises if an electric current $\boldsymbol{j_c}$ is sent along one of these directions and the current is spin-polarized due to difference in the Fermi surfaces for spin up and down [Fig. 1(d)]. However, the spin-splitter effect appears when $\boldsymbol{j_c}$ is applied along a nodal direction [100] or [010] and a transversal spin current $\boldsymbol{j_s}$ arises as a consequence of anisotropic spin-splitting [Fig. 1(e)], leading to spin-dependent deflection of electrons. The inverse analog of the latter phenomenon is dubbed ISSE. The spin-to-charge current conversion due to it is represented by the angle $\gamma'$ and is added to the ISHE in the emission scheme described above [Fig. 1(f)].

Since the direction of the spin-to-charge current conversion due to the ISSE is determined by the fixed crystal orientation, the emission texture is anisotropic. If we set $\theta = 0°$ for $\boldsymbol{M} \parallel \mathrm{y}$, i.e., the $[001]$ direction of $RuO_2$, we expect only the $y$-projection of the injected spin to contribute to $E_x$ of the emission texture and its $\theta$-dependence as shown in Fig. 1(g, h). Notably, $E_x \propto \cos\theta$, $E_y = 0$ and $E_r \propto |\cos\theta|$, which allows us to disentangle the isotropic ISHE and the anisotropic ISSE. From these considerations, we can formulate two basic criteria to confirm the presence of the ISSE in $RuO_2$-based samples: (1) The THz emission from $RuO_2$(100) must be consistent with the texture in Fig. 1(g). (2) This contribution must vanish in RuO(110), where no ISSE signal is theoretically expected from symmetry, Fig. 1(e) [2,10] and used for interpretation of experiments [11,12,14].

These criteria dictate the choice of our samples: 10-nm-thick (100) or (110)-oriented $RuO_2$ grown on a (100) or (110)-oriented $TiO_2$ substrate, followed by 2.5 nm of ferromagnetic $Co_{40}Fe_{40}B_{20}$, capped by 3 nm of Si (see Supplementary Note 1 for more sample characterization) [34]. As reference sample, we use the same FM layer grown on $TiO_2$(100), capped by 2.5 nm of Pt. The in-plane orientation of the CoFeB magnetization is controlled by a rotatable external magnetic field of strength of 350mT, which is more than an order of magnitude larger than the saturation field of the FM layers (see Supplementary Fig S5) [34]. To excite the thin-film stacks, we use femtosecond laser pulses generated in a regenerative amplifier (pulse duration 170 fs, energy 30 μJ, repetition rate 10 kHz). The emitted THz field components $E_x$ and $E_y$ are probed by electro-optic (EO) sampling [35,36] in a 2-mm-thick GaP(110) detection crystal using two THz polarizers, resulting in EO signals $S_x^{EO}$ and $S_y^{EO}$. All experiments are performed in a dry-air atmosphere and at room temperature.

The typical $\boldsymbol{M}$-dependent THz waveforms $S_x\,(t)$ and $S_y\,(t)$ from our samples, after subtraction of the $\theta$-averaged, nonmagnetic signal from $S_x^{\mathrm{EO}}$ and $S_y^{\mathrm{EO}}$, are shown in Fig. 2(a) and (b). While the signal from the CoFeB|Pt reference sample (magenta curves) shares practically identical temporal dynamics in both $x$- and $y$-projections, the signal from $RuO_2$|CoFeB (red and blue curves) shows a notable difference. Mainly, $S_y(t)$ has a more bipolar character whereas $S_x(t)$ is more unipolar, similar to signals from CoFeB|Pt. This behavior suggests that the spin-to-charge conversion process may partly have a different origin. The full $\theta$-dependence of signals is shown for (100)-RuO2|CoFeB by a sequence of 40 colors.

Further, we measure the $\theta$-dependence of $S(t)$ for all samples (shown in Fig. 2 by sequence of 40 colors and S1) [34], compute the root-mean-square (RMS) of the waveforms, multiplied by the sign of the signal at the main waveform peak, and plot them in the form of emission textures [Fig. 3(a, c, e)] and of the corresponding $S_x(\theta)$, $S_y(\theta)$ and $S_r(\theta)$ projections [Fig. 3(b, d, f)]. We note that RMS is a convenient way to determine the signal strength as it less prone to single-datapoint noise as compared to peak-to-peak value. Without any quantitative analysis, we see that all the emission textures are not fully isotropic, as expected from the purely isotropic-ISHE-driven conversion [cf. Fig. 1(b, c)]. To highlight the differences, we model the isotropic contribution by fitting $S_y$ by a harmonic function, $S'_y = -A_i \sin\theta$, where $A_i$ with $i = Pt, (110), (100)$ is the amplitude of the isotropic conversion for the reference CoFeB|Pt and the respective orientation of $RuO_2$|CoFeB samples. Subsequently, we plot the corresponding $x$-projection, $S'_x = A_i \cos\theta$, and the magnitude of the emission $S'_r = A_i$ [curves in Fig. 3(b, d, f)], and shade the difference between the model and the experimental data. The best-fit parameters $A_{(100)} = 0.024 A_{\mathrm{Pt}}$ and $A_{(110)} = 0.031 A_{\mathrm{Pt}}$ allow us to estimate the effective spin Hall angle $\gamma_{\mathrm{RuO2}}$ of the isotropic conversion in $RuO_2$ by a comparison to $A_{Pt}$ of the reference sample. By assuming that the total conversion is dominated by the conversion inside $RuO_2$ and $\gamma_{Pt} = 10^{-1}$ [37,38,39,40,41] (see Supplementary Note 2 for more details) [34], we obtain $\gamma_{\mathrm{RuO_2}} = -(2.1 \pm 0.2) \cdot 10^{-3}$ and $-(2.7 \pm 0.2) \cdot 10^{-3}$ for the (100) and (110)-oriented sample, respectively. Note that the negative sign of $\gamma_{RuO2}$ comes from the fact that the signal polarity is the same as for CoFeB|Pt (Fig. 2), but the layer order of the FM|HM stacks is reversed (Fig. 1).

The additional conversion contributions are analyzed by subtraction of the isotropic model from the data, i.e., $\Delta S_x = S_x - S'_x$ and $\Delta S_y = S_y - S'_y$, and by normalizing them to the corresponding $A_i$, as shown in Fig. 4. The residual emission textures (panels a, c, e) and their projections (panels b, d, f) follow an ISSE-like pattern [Fig. 1(d, f)] in all three samples. Remarkably, the modulation depth of $\Delta S_x / A_i$, which quantifies the anisotropic emission, reaches $(46 \pm 1)\%$ and $(53 \pm 1)\%$ for (100)-$RuO_2$ and (110)-$RuO_2$, respectively, and $(28.8 \pm 0.3)\%$ for the reference sample. The very similar modulation strengths in both orientations (100) and (110) of $RuO_2$ clearly violate the criterion (2). The small difference in the modulations indicates that a possible contribution of the ISSE remains very limited in our measurements, with an upper bound of several percent of the magnitude of the isotropic ISHE conversion.

Next, we investigate the sources of anisotropies in emission that may appear in our samples. First, the used $TiO_2$ substrate exhibits a very strong birefringence in the THz spectral range [42,43], which affects the outcoupling of the THz radiation from the sample. As anticipated by Eq. (1), the outcoupling is characterized by the sample impedance $Z$, which can be modeled in the thin-film approximation [29,44] as $Z \approx Z_0/(1 + n_s + Z_0 G)$. Here, $Z_0 \approx 377\Omega$ is the vacuum impedance, $n_s$ the refractive index of the substrate in the polarization direction of $E$, and $G$ is the THz conductance of the conductive layers. We directly measured the two principal values of $n_s$ in uniaxial $TiO_2$ substrates in the THz range (see Supplementary Fig. S2) [34], showing no significant spectral dependence and yielding $n_{s,x} = 9.5 \pm 0.1$ and $n_{s,y} = 12.9 \pm 0.1$, respectively. The substrate birefringence leads to an anisotropic impedance: $Z_x = Z_y + \delta Z$, and accounts for its notable relative variation $\delta Z / Z_y = (26 \pm 1)\%$ for average $G$ of our samples.

Another source of anisotropic outcoupling could be the anisotropic $G$ of the conductive layer itself, which was suspected to cause the anisotropic THz emission from $RuO_2$-based samples in Ref. 27. However, using THz transmission experiments, we did not observe any large variation of $G$ in our samples as $\delta G / G_y \approx 1.4\%$ and $-4.3\%$ for (100)- and (110)-$RuO_2$|CoFeB, respectively, while in the case of CoFeB|Pt we inferred $\delta G / G_y \approx 18.6\%$ (see Supplementary Table T1) [34].

The third potential source is the recently suggested anisotropy of the spin Hall angle in (100) and (110)-oriented $RuO_2$, which was evaluated by DC spin Seebeck experiments to $\delta\gamma/\gamma_y \approx 30\%$ [25]. Using the above-mentioned anisotropies, we can model the anisotropic emission by separating the $x$- and $y$-components of the emitted $E$ and expanding Eq. (1) as follows (see Supplementary Note 3):

$$\begin{pmatrix} E_x \\ E_y \end{pmatrix} = e \begin{pmatrix} Z_x \boldsymbol{I}_{c,x} \\ Z_y \boldsymbol{I}_{c,y} \end{pmatrix} \propto \begin{pmatrix} Z_x(\gamma_x + \gamma') \cos\theta \\ Z_y \gamma_y (-\sin\theta) \end{pmatrix} = Z_y \gamma_y \left[ \begin{pmatrix} \cos\theta \\ -\sin\theta \end{pmatrix} + \begin{pmatrix} \delta\gamma/\gamma_y + \gamma'/\gamma_y + \delta Z/Z_y + \Delta \\ 0 \end{pmatrix} \cos\theta \right] \quad (2)$$

Here, both the $x$- and $y$-components include their respective outcoupling $Z$, the spin Hall angle $\gamma$ is anisotropic and $\gamma'$ contributes only to $E_x$. The first term after the last equality is the isotropic conversion due to the isotropic ISHE, the second one is the correction due to anisotropies and possible ISSE, while $\Delta = (\delta Z/Z_y)(\delta\gamma/\gamma_y) + (\delta Z/Z_y)(\gamma'/\gamma_y)$ are less significant second order terms. We see from Eq. (2) that all anisotropies contribute to the correction with the identical symmetry as the ISSE and, therefore, this altermagnetic feature cannot be separated based solely on the $\theta$-dependence of the THz emission from one sample.

Therefore, we can compare the THz emission from all three samples with the model by plotting the anisotropic term of Eq. (2) on top of the measured data in Fig. 4(b, d, f). First, we consider only the outcoupling anisotropy $\delta Z/Z_y$ and assume $\gamma' = \delta\gamma = 0$ (dashed violet curves). We observe a perfect match in the case of the reference sample, CoFeB|Pt, which is consistent with the expectation that Pt does not show any anisotropic ISHE. On the other hand, the data from the $RuO_2$-based samples clearly cannot be explained just by the outcoupling anisotropies (i.e., $\delta n_s$ and $\delta G$). However, once we also add the anisotropic ISHE and fit the modulation of (110)-$RuO_2$ (full orange curves), we obtain $\delta\gamma/\gamma_y = (23 \pm 2)\%$. This value is consistent with the recent experimental observation [25]. By considering the anisotropy $\delta\gamma/\gamma_y$ to be same in both $RuO_2$ samples, we use $\delta\gamma/\gamma_y$ obtained from (110)-$RuO_2$ also to model the response of (100)-$RuO_2$ (labeled Model 1). Here, we see a very good match of the model with the measured data, showing a difference by $\sim 9\%$. From this observation and from the fitted isotropic $\gamma_y \sim 2.4 \cdot 10^{-3}$, we can infer the upper bound of the magnitude of the potential, area-averaged contribution of the ISSE to $\gamma' \approx 2 \cdot 10^{-4}$ in our (100)-$RuO_2$ sample at room temperature. This is three orders of magnitude lower value than the theoretical prediction of the zero-temperature ISSE in the single-domain altermagnetic phase of $RuO_2$ [2,10]. In addition, the amplitude of the anisotropic emission from $RuO_2$-based samples tended to decrease in time, and it shows no observable change when heated by 40°C above the room temperature (Supplementary Fig. S3) [34].

The approximate isotropic $\gamma_{RuO2} = -(2.4 \pm 0.2) \cdot 10^{-3}$, averaged over both orientations of our $RuO_2$ samples, deviates from values reported in previous experiments in Refs. [11,12,13,23,26], where it had the opposite sign and ranged from $1.5 \cdot 10^{-2}$ to $4.5 \cdot 10^{-2}$. However, the polarity of our inferred SHE angle matches the recent study findings by Wang et al. in YIG/RuO2 systems [25]. This work was performed on a large set of samples with an insulating ferrimagnetic layer, thus, without an additional contribution from the spin-to-change conversion in the magnet. The authors assign the overall positive polarity of the previous studies to this potential contribution of the positive SHE angle in metallic metals like NiFe and CoFeB [45,46]. Considering the generally lower $\gamma_{FM}$ in $Co_{40}Fe_{40}B_{20}$ [46], it may explain our lower and negative value of $\gamma$. Other possible ISHE-like mechanisms contributing to the overall emission, originating in the ferromagnet, are the ultrafast demagnetization [32], the anomalous Hall effect [47] and anomalous Nernst effect [48]. All these mechanisms, summarized in an effective conversion angle $\gamma_{FM}$, are expected to be relatively weak in thin FM layers as compared to the leading mechanism, ISHE from $RuO_2$, and, more importantly, fully isotropic, as they are originating from the isotropic polycrystalline CoFeB layer. Therefore, they could not contribute to the anisotropic parts of the conversion, the altermagnetic $\gamma'$ or the anisotropic ISHE $\delta\gamma$. However, they can, in principle, contribute to the isotropic ISHE conversion $\gamma_y$ itself. Additionally, the conversion by ISHE inside $RuO_2$ can be impacted by spin losses at the FM/$RuO_2$ interface [37], resulting in its lower spin transparency $t_s$, or the Rashba-like conversion $\gamma_I$ at the interface itself.

These probably small, but potentially present contributions are generally difficult to be disentangled in THz emission experiments [37,49], and we advise to consider the isotropic part of $\gamma$ the total effective conversion efficiency of the particular stack of layers:

$$\gamma_y = \gamma_{FM} + \gamma_I + t_s \gamma_{RuO2} \quad (3)$$

In Fig. 3 (e) and (f), we see that the total isotropic conversion $\gamma_y$ (i.e., the amplitude of $S_y$) varies by ~20% between the sample stack with (110)-$RuO_2$ and (100)-$RuO_2$. In the above estimate of the size the altermagnetic conversion $\gamma'$, we used the relative value of the anisotropic ISHE, $\delta\gamma/\gamma_y$, from (110)-$RuO_2$ sample in the model of emission from (100)-$RuO_2$ (Model 1). This approach is immune to the variation of $\gamma_y$ due to spin loss at the interface, which we consider the probable source of it, as $t_s$ cancels in the term $\delta\gamma/\gamma_y$. However, we can explore the other extremal case when the difference of the isotropic $\gamma_y$ is solely due to a change in $\gamma_{FM}$ and $\gamma_I$, implying a variation of the CoFeB layer in these two stacks. To eliminate this effect from the analysis, one needs to use only δγ from (110)-$RuO_2$ sample to model the conversion in (110)-$RuO_2$. This hypothetical situation leads to a proportionally larger modulation [Model 2, see red curve in Fig. 4(f)] and yielding the maximal ISSE contribution $\gamma' \approx 4 \cdot 10^{-4}$. The realistic model, therefore, lies between Model 1 and 2, giving the upper bound of γ′ in the interval $(2-4) \cdot 10^{-4}$.

The absence of a sizable ISSE in our samples at room temperature is the central conclusion of our paper. It is consistent with its absence reported by the very recent DC thermo-transport [25] and THz emission experiments [27], and contrasts with the earlier SSE spin pumping studies at room temperature [11,12,13], where the authors inferred a magnitude of the SSE conversion in the same order as the SHE. Remarkably, experiments in Ref. [11] showed a strong temperature dependence of the SSE, which also contrasts with the negligible temperature dependence of our anisotropic signals (Supplementary Fig. S3) [34]. Following the ongoing intense debate on the magnetic nature of $RuO_2$, there are two main explanations of our reported the potential weak ISSE: 1) The magnetic domain structure can be formed with mutually compensating contributions to the ISSE. This argument was already used in Refs. [11] and [13] when explaining why the inferred ISSE reached only a fraction of the theoretically expected value [10]. 2) The debated non-magnetic nature of $RuO_2$ at room temperature in bulk and thin layers [16,17,18]. Reflecting on the ongoing discussion of the magnetic order, we stress the relevance of our study with new insights in origin of anisotropic signals in $RuO_2$ and we suggest that future studies should include the THz emission experiments at lower temperatures [19,20] and in $RuO_2$ films of a variable disorder/impurity level [50] and thickness, including ultrathin, strained $RuO_2$ films with layer thicknesses below 2 nm [21,51]. The last-mentioned ultrathin strained films might be particularly interesting for the THz emission experiments, as this technique is interface sensitive and was successfully applied to nanometer-thin layers [37,49].

In summary, we performed THz-emission spectroscopy on $RuO_2$|CoFeB bilayers and studied the phenomenology of the spin-to-charge current conversion, with a particular focus on the observed anisotropic emission. To carefully analyze the data, we modeled the emission by considering the predicted altermagnetic contribution of the ISSE, together with three additional sources of analogous anisotropy: the birefringence of the substrate, the anisotropic ISHE and the anisotropic conductivity of $RuO_2$. The modeling revealed that the anisotropic ISHE, recently proposed [26,25] due to the crystalline symmetry of $RuO_2$ and unprecedented in THz experiments, is indeed operative in the THz spectral range and that it reaches similarly large magnitude in both (100)- and (110)-grown $RuO_2$. This finding can have an application potential in devices which employ the transient spin accumulation or spin transfer torque, where the current direction can change the magnitude of the effect. Importantly, we observed no significant contribution of the altermagnetic ISSE at room temperature with a magnitude $(2-4) \cdot 10^{-4}$. Our qualitative and quantitative understanding of the interplay of all sources of anisotropic THz emission is an important ingredient for the analysis and designs of future optical and THz experiments in $RuO_2$.

**Acknowledgements**

The authors acknowledge funding by the Czech Science Foundation through projects GA CR (Grant No. 25–18281K), the Grant Agency of the Charles University (SVV–2024–260720), the Ministry of Education, Youth and Sports of the Czech Republic through the OP JAK call Excellent Research (TERAFIT Project No. CZ.02.01.01/00/22_008/0004594), the ERC-2023 Advanced Grant ORBITERA (grant No. 101142285), the ERC-2023 Advanced Grant ALTERMAG (grant No. 101095925), the ERC-2023 Proof-of-Concept Grant T-SPINDEX (grant No. 101123255), the DFG Collaborative Research Center SFB TRR 227 "Ultrafast spin dynamics" (project ID 328545488, projects A05 and B02), the DFG Excellence Cluster EXC 3112 "Center for Chiral Electronics" (EXC 3112/1, Project No. 533767171) and the Research Unit RU 5844 ChiPS (Project No. 541503763). S.M and P.Ke. received funding from the Deutsche Forschungsgemeinschaft (DFG, German Research Foundation) under Germany's Excellence

Strategy through the Würzburg-Dresden Cluster of Excellence on Complexity and Topology in Quantum Matter ct.qmat (EXC 2147, Project ID 390858490) and through the Collaborative Research Center SFB 1170 ToCoTronics (Project ID 258499086), as well as from the New Zealand Ministry of Business, Innovation and Employment (MBIE, Grant number: C05X2004). G.D.L. acknowledges Grant No. RYC2021-032524-I funded by MCIN/AEI/10.13039/501100011033 and by European Union NextGenerationEU/Plan de Recuperación, Transformación y Resiliencia. H.R. to the Dioscuri Program LV23025 funded by Max Planck Society and MEYS of the Czech Republic. J. J. acknowledges the support by the Grant Agency of the Charles University (Grant No. 120324). P. Ku. acknowledges the support by the Grant Agency of the Charles University (Grant No. 166123).

*Data availability*—The data that support the findings of this Letter are openly available [52].

---

Supplementary material for

# Spin- to charge-current conversion in altermagnetic candidate $RuO_2$ probed by terahertz emission spectroscopy

J. Jechumtál[1], O. Gueckstock[2], K. Jasenský[1], Z. Kašpar[1,3], K Olejník[3], M. Gaerner[4], G. Reiss[4], S. Moser[5], P. Kessler[5], G. De Luca[6], S. Ganguly[7], J. Santiso[7], D. Scheffler[3], J. Zázvorka[1], P. Kubaščík[1], H. Reichlova[3], E. Schmoranzerova[1], P. Němec[1], T. Jungwirth[3,8], P. Kužel[3], T. Kampfrath[2] and L. Nádvorník[1*]

1. Faculty of Mathematics and Physics, Charles University, 121 16 Prague, Czech Republic
2. Department of Physics, Freie Universität Berlin, 14195 Berlin, Germany
3. Institute of Physics, Czech Academy of Science, 16200 Prague, Czech Republic
4. Bielefeld University, Faculty of Physics, 33615 Bielefeld, Germany
5. Physikalisches Institut and Würzburg-Dresden Cluster of Excellence ct.qmat, Universität Würzburg, 97074 Würzburg, Germany
6. Materials Science Institute of Barcelona (ICMAB-CSIC), 08193 Barcelona, Spain
7. Catalan Institute of Nanoscience and Nanotechnology, 08193 Barcelona, Spain
8. School of Physics and Astronomy, University of Nottingham, UK

* E-mail: lukas.nadvornik@matfyz.cuni.cz

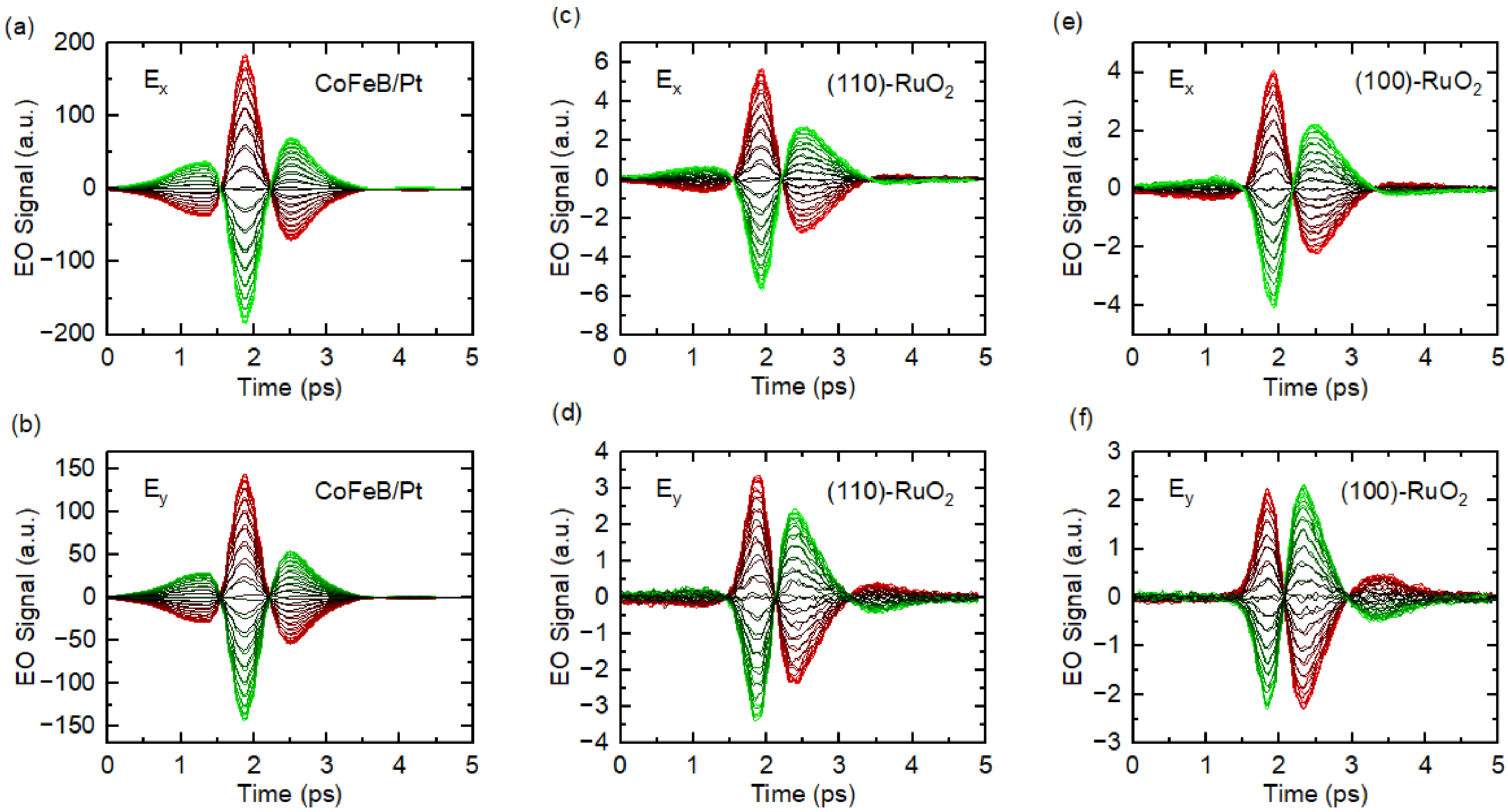

**Figure S1 | Full θ-dependent signals $S_x$ and $S_y$ measured on CoFeB/Pt, (110)-$RuO_2$ and (100)-$RuO_2$.** The $x$ and $y$ projections of the EO signals $\boldsymbol{S}(\boldsymbol{t})$ from CoFeB/Pt (a) and (b), (110)-RuO2/CoFeB (c) and (d) and (100)-RuO2/CoFeB (e) and (f). The full $\boldsymbol{\theta}$-dependence of signals is shown by a sequence of 40 colors, same as in the main text.

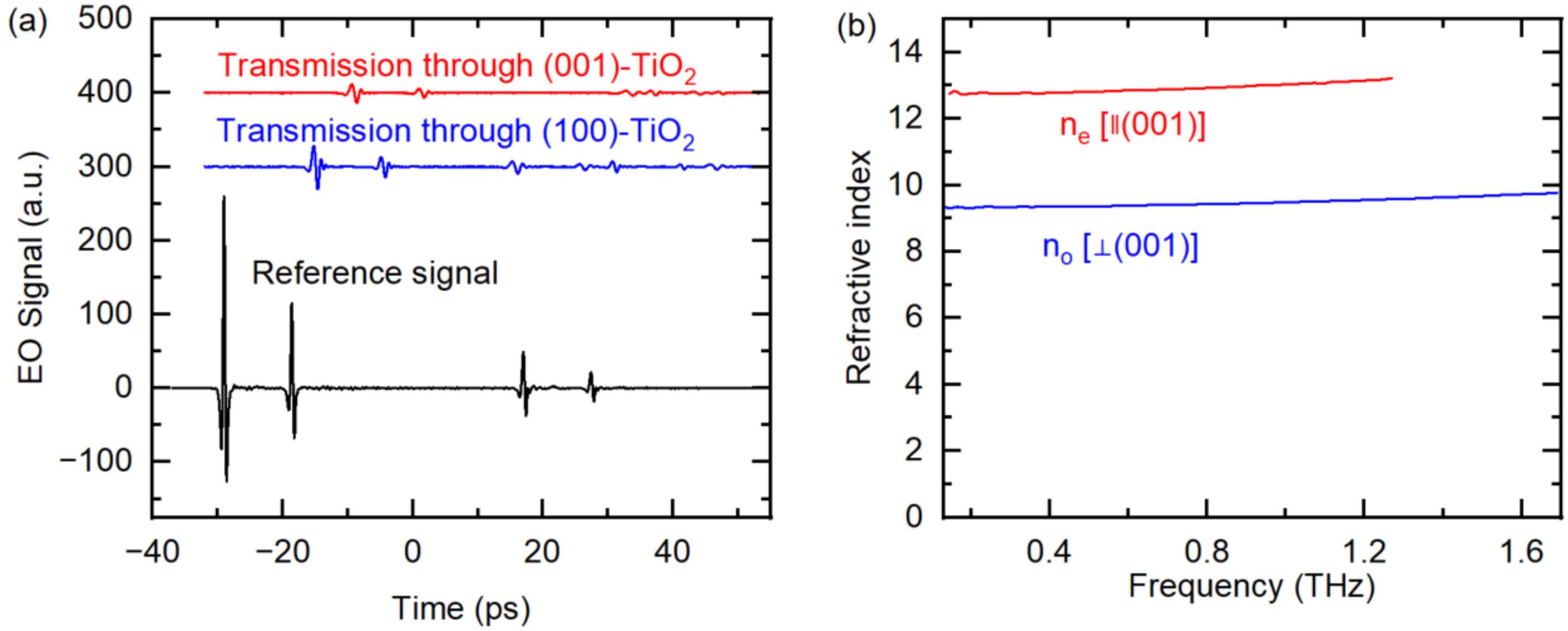

**Figure S2 | Ordinary and extraordinary refractive index of $TiO_2$**. Refractive indices of $TiO_2$ shown in panel (b) were extracted from experimentally measured THz transmission waveforms shown in panel (a) [1]. Measured substrates were from the same batch as the substrates used for the sample growth. The imaginary part of the refractive index is negligible.

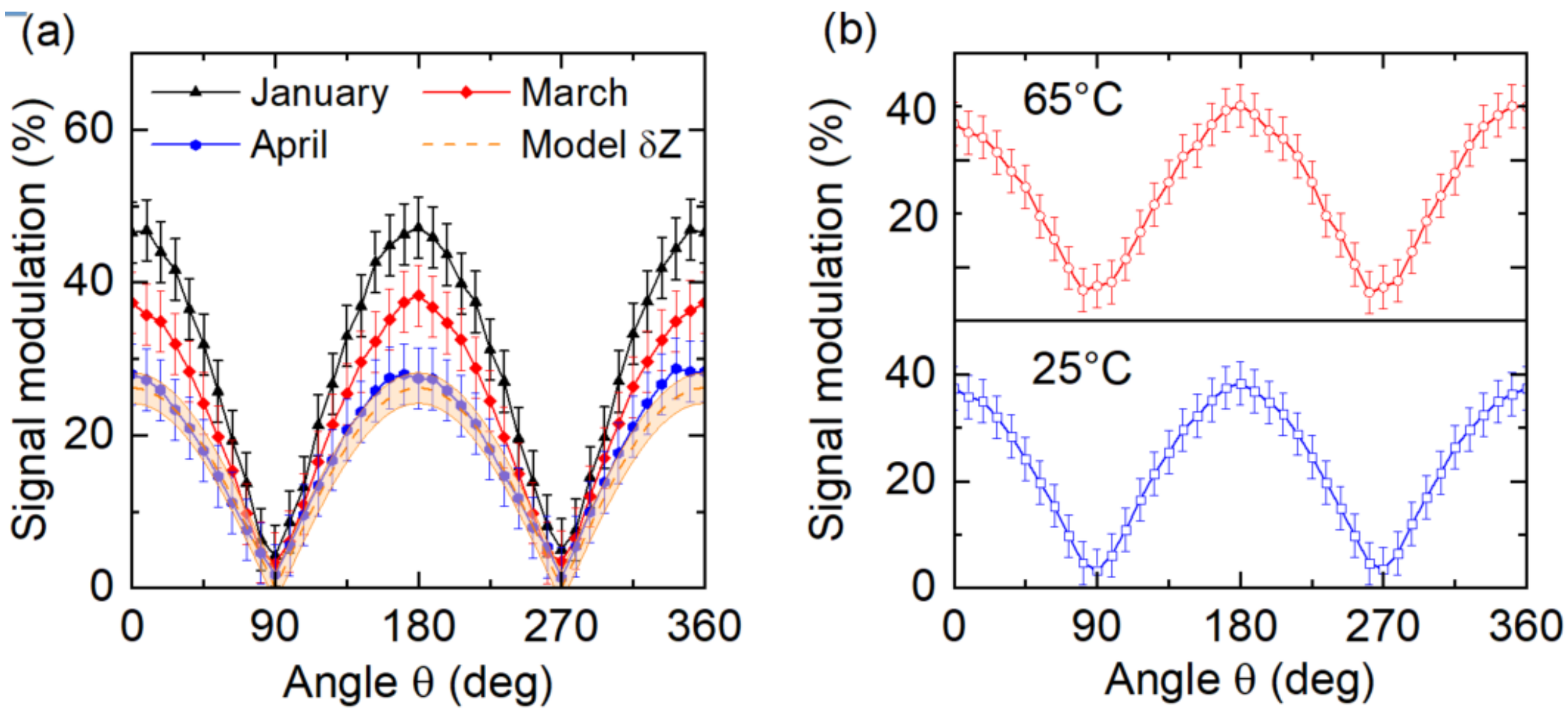

**Figure S3 | Time and temperature dependence of emission in (100)-RuO2|CoFeB**. (a) The $\boldsymbol{\theta}$-dependence of signal modulation $\boldsymbol{\Delta S_r}$ as a function of date of measurement. The orange dashed curve is the model considering just the birefringence of the substrate (only $\boldsymbol{\delta Z/Z \neq 0}$). Over four months (measurement done always at the end of the given month, approximately 2 months after sample fabrication) $\boldsymbol{\Delta S_r}$ decayed almost to this residual anisotropic outcoupling effect, indicating that the anisotropic ISHE component is diminishing in macroscopic time. (b) The $\boldsymbol{\theta}$-dependence at the room and elevated temperature, showing no observable reduction of the modulation. This measurement was done in March.

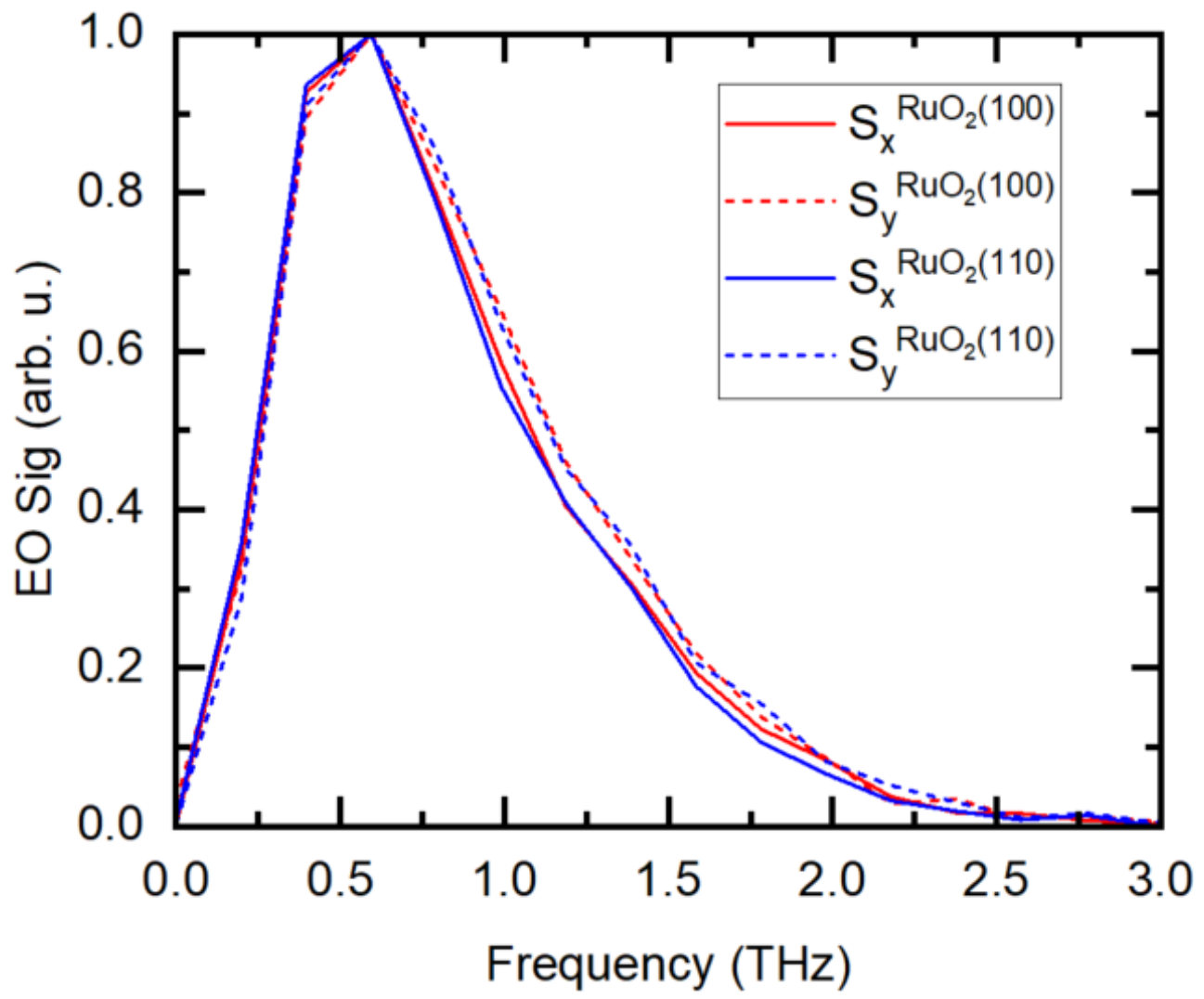

**Figure S4 | Spectra of THz emission from $RuO_2$|CoFeB samples**. Normalized spectra of magnetic signals from $RuO_2$|CoFeB samples show the same behavior of (100)- and (110)- $RuO_2$ sample and small low-pass filtering of $S_x$ when compared to $S_y$, which is explainable by different outcoupling conditions caused by anisotropic $TiO_2$ substrate.

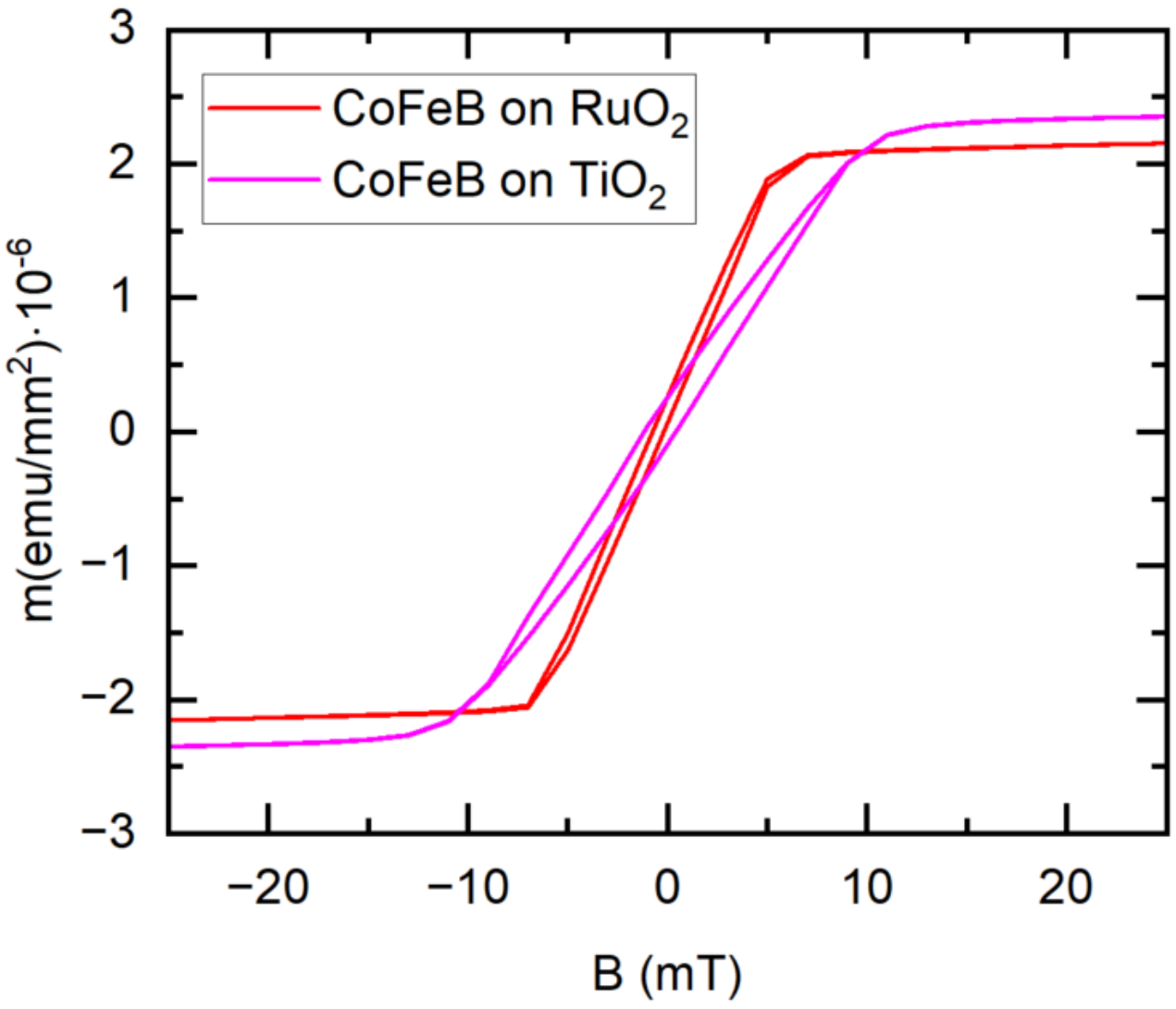

**Figure S5 | SQUID characterization of CoFeB layer grown on $TiO_2$ and $RuO_2$**. The magnetization measurement of two CoFeB layers normalized on the sample area show very similar behavior and full saturation above 12 mT. The slightly smaller value of magnetic moment in CoFeB on $RuO_2$ is explainable by the error of estimation of the exact area of measured sample.

| | $RuO_2(100)$/CoFeB | $RuO_2(110)$/CoFeB | Reference STE on $TiO_2(100)$ |
|---|---|---|---|
| $G_1$ (mS) | $G_{001}$ 7.1 +- 0.4 | $G_{001}$ 6.7 +- 1 | $G_{001}$ 7.0 +- 0.3 |
| $G_2$ (mS) | $G_{010}$ 7.0 +- 0.6 | $G_{1-10}$ 7.0 +- 1 | $G_{010}$ 5.9 +- 0.6 |
| Expected modulation (%) | 26 +- 1 | 25 +- 1 | 30.0 +- 1 |

**Table T1 | Conductances and expected signal modulation from anisotropic $\boldsymbol{Z}$ and $\boldsymbol{G}$.** Conductivities in two parallel directions of the used samples were measured via THz transmission spectroscopy. The mean value in the usable spectral range is shown together with the standard deviation. The expected 180°-modulation of the THz signal, caused by the anisotropic impedance $Z$, is shown in the 3rd row.

# Note 1 | Sample characterization

The samples were grown by magnetron sputter deposition in a UHV chamber with a base pressure $< 1 \cdot 10^{-9}$ mbar. The $RuO_2$(100) and (110) layers with nominal thicknesses of 10 nm were grown simultaneously via reactive RF-sputtering onto $TiO_2$(100) und (110) substrates with an Ar:$O_2$ ratio of 4:1 and at a substrate temperature of about 230°C. The samples were then cooled down to room temperature. Subsequently, a $Co_{40}Fe_{40}B_{20}$ layer of 2.5 nm nominal thickness was deposited on top of the $RuO_2$ via DC-sputtering. The samples were capped by an RF-sputtered Si layer of 3 nm nominal thickness. Prior to the deposition of both $Co_{40}Fe_{40}B_{20}$ and Si, the targets have been pre-sputtered for about 20 minutes to get rid of oxides on the target surfaces which could have formed during the reactive sputtering process.

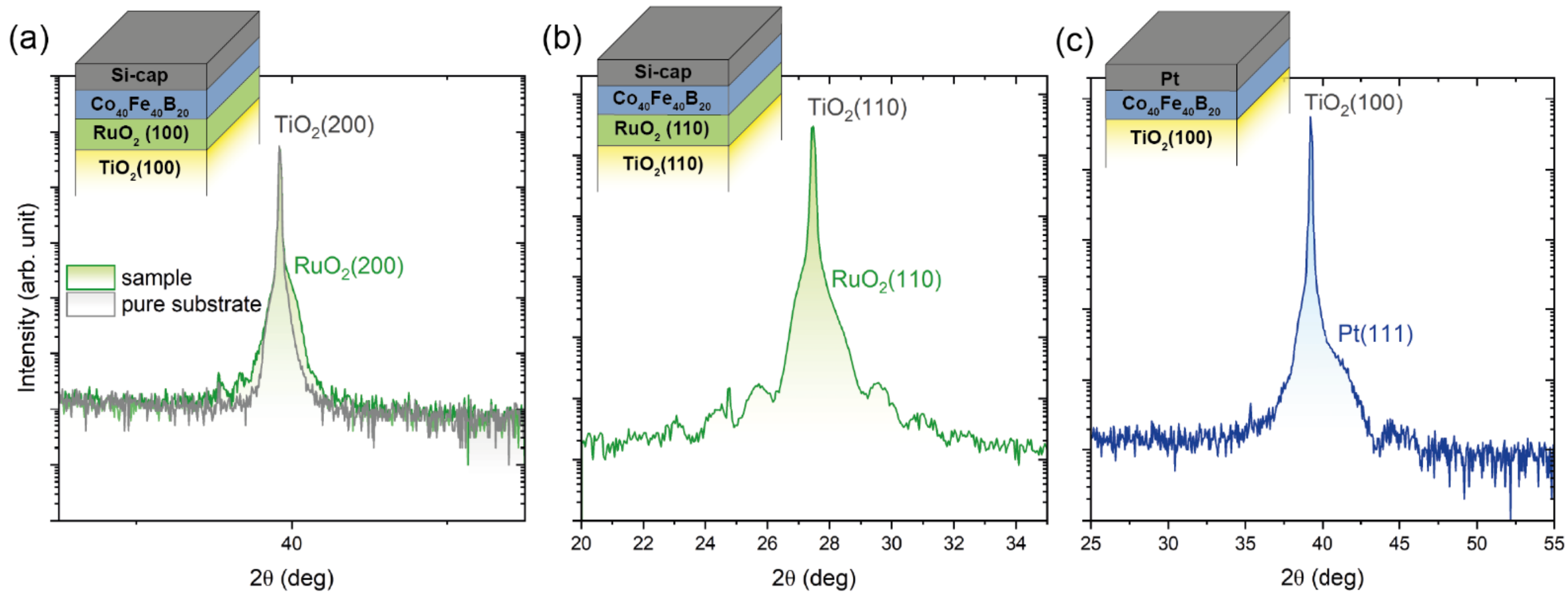

**Figure S4 | XRD sample characterization.** (a) Measurement of the $RuO_2$(100) sample as well as a pure substrate for reference, (b) the $RuO_2$(110) sample and (c) the reference sample with Pt capping. The exact layer structures are shown in schematics.

The prepared samples were characterized by specular X-ray diffraction (XRD) and X-ray reflectivity (XRR) measurements using the Cu Kα ($\lambda = 1.54056$ Å) source of a PANalytical X'Pert Pro MPD PW3040-60 diffractometer. As shown in Figs. S1(a-c), the XRD peaks of the thin films overlap with the substrate peaks in all cases, as $TiO_2$ and $RuO_2$ can be considered isostructural and possess similar lattice constants. However, Laue oscillations can clearly be observed for the $RuO_2$(110) sample but are also recognizable in the other two samples, which indicates good crystalline quality of the $RuO_2$ layers as well as the Pt layer. The XRD scans have been repeated after the THz characterization. However, no crystallographic changes were noticeable.

## Note 2 | Details on the evaluation of the spin Hall angle in $RuO_2$

For an evaluation of the Spin Hall angle in $RuO_2$, we compare the amplitude of isotropic THz emission in the sample containing 10 nm thick $RuO_2$ with the reference one consisting of a conventional FM/Pt bilayer; both structures are shown in Fig. 1 in the main text. In the following text we use the subscripts $RuO_2$ and Pt to refer to the quantities related to the sample with $RuO_2$ and to the reference sample, respectively.

Following Eq. (1) in the main text we can write for the ratio of the spin Hall angles:

$$\frac{\gamma_{\mathrm{RuO_2}}}{\gamma_{\mathrm{Pt}}} = \frac{S_{\mathrm{RuO_2}}}{S_{\mathrm{Pt}}} \frac{Z_{\mathrm{Pt}}}{Z_{\mathrm{RuO_2}}} \frac{P_{\mathrm{Pt}}}{P_{\mathrm{RuO_2}}}. \quad \text{(S1)}$$

Here $\gamma$ is the spin Hall angle, $S$ is the magnitude of the measured emitted THz field, $Z$ is the overall impedance of the sample [1] and $P$ is the energy density absorbed in the FM CoFeB layer (i.e., we assume that the magnitude of the induced spin current $j_s$ is proportional to the energy deposited by the pump laser pulse in the FM layer).

The conductances of the samples as measured by terahertz transmission spectroscopy (Table T1), are very similar, (100)-$RuO_2$/CoFeB: 7.1 mS, (110)-$RuO_2$/CoFeB: 6.7 mS, CoFeB/Pt: 7 mS; thus, the difference in the impedance in the given in-plane $TiO_2$ crystal orientation (001) of the samples can be neglected: $Z_{\mathrm{Pt}}/Z_{\mathrm{RuO_2}} \approx 1$.

It is then necessary to evaluate carefully the energy deposited by the laser pulse in the CoFeB layer, which serves as the spin current source. For this calculation, we employ the transfer matrix formalism [2,3], which allows us to calculate the optical pump field distribution $E_o(z)$ inside the CoFeB layer in both samples. Since the magnetic properties of CoFeB vanish at optical frequencies ($\mu(\omega_o) = 1$), the absorbed optical energy density is then equal to its dielectric part [4]:

$$P(z) = \frac{1}{2}\omega_o \, Im(\varepsilon) \, |E_o(z)|^2 \equiv \alpha \, I(z) \quad \text{(S2)}$$

where $I(z)$ is the z-dependent power density of the optical beam in CoFeB and $\alpha$ is the absorption coefficient of CoFeB. The total absorbed energy density is then equal to

$$P = \int_0^d P(z), \quad \text{(S3)}$$

where $d$ is the thickness of CoFeB layer.

We use the following values of complex refractive indices $N = n + i\kappa$ at the excitation wavelength of 1030 nm, $TiO_2$: $N = 2.48$ [5], CoFeB: $N = 4 + i\,4.7$ [6], $RuO_2$: $N = 1.2 + i\,2.3$ [7], Pt: $N = 0.844 + i\,10.6$ [8], Si: $N = 3.565 + i\,0.00024$ [9].

Since the $TiO_2$ substrate is thick, the pulse train created by multiple internal reflections (etalon effect) of the pump pulse in the substrate is well separated in time from the part of the pump pulse directly passing through the substrate. Indeed, our THz emission experiments are related only to this direct-pass pump. In our calculation of the absorbed energy density in the CoFeB layer we thus considered only the direct pass in the substrate and all the multiple reflections in the thin film sequence. In Fig. S5, we show the obtained profile of $|E_o(z)|^2$ in both the sample and reference structures: this yields the value $P_{\mathrm{Pt}}/P_{\mathrm{RuO_2}} \approx 0.9$.

Please note that in $TiO_2$ (yellow zone in Fig. S5), we observe an interference of the incident and reflected waves. The physical reason for the small difference between the energy absorbed in the FM layer in the two structures consists essentially of a different impedance mismatch between the substrate and the layered structure in the two cases. This mismatch controls the phase of the reflected wave at this interface and, consequently, the overall level of the optical power density in the layered structure.

To check the plausibility of our findings, we measured the power reflectance and transmittance of the sample and of the reference structures illuminated by an incident beam at 1030 nm; subsequently, we compared these values to the calculated ones using our transfer matrix model supplied with the values of complex refractive indices provided above. (Note that this transmission and reflection experiment is not time resolved and thus it involves all the multiple reflections in the thick $TiO_2$ substrate; our calculations then properly took into account these multiple reflections). The comparison of the theoretical and experimental values is shown in Table T2; a reasonable agreement is found both for the sample and the reference structure.

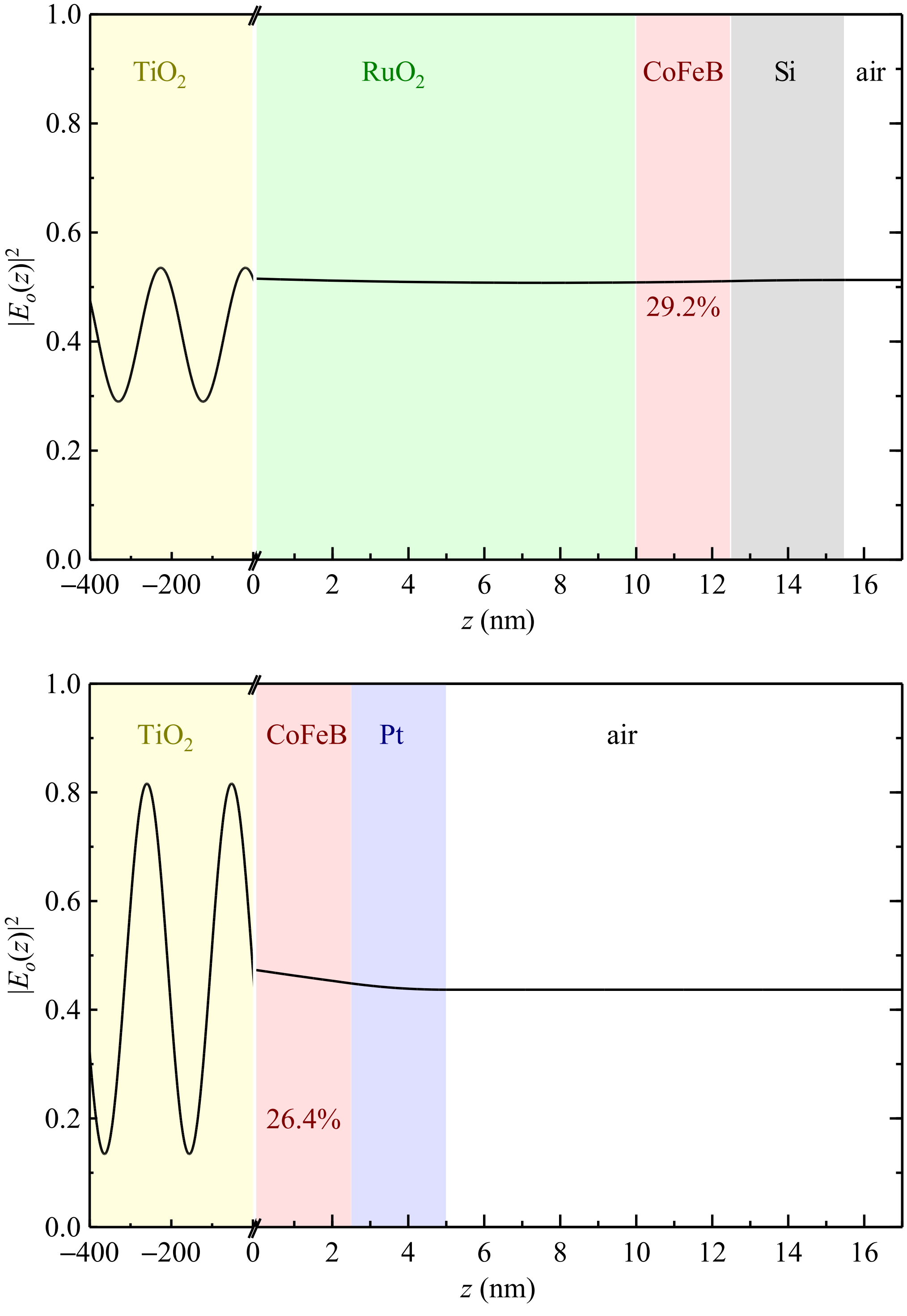

Fig. S5: Distribution of $|E_o(z)|^2$ in the samples upon optical excitation at 1030 nm. Top panel: sample structure; bottom panel: reference structure. The amount of optical power density absorbed in the CoFeB film is also indicated.

Table T2: Power transmittance and reflectance of the sample and reference structures at 1030 nm; comparison of the model and experimental results.

| Structure | Transfer matrix model | Experiment |
|---|---|---|
| $TiO_2$ / CoFeB / Pt (reference) | $T = 37\ \%, R = 30\%$ | $T \approx 35\ \%, R \approx 20\%$ |
| $TiO_2$ / $RuO_2$ / CoFeB / Si (sample) | $T = 42\ \%, R = 20\%$ | $T \approx 40\ \%, R \approx 20\%$ |

## Note 3 | Emission model of $RuO_2$ (100)-based THz emitters

To derive equation (2) in the main text, we use the description of the THz field vector from the main text and expand:

$$\begin{pmatrix} E_x \\ E_y \end{pmatrix} = e \begin{pmatrix} Z_x I_{c,x} \\ Z_y I_{c,y} \end{pmatrix} \propto \begin{pmatrix} Z_x(\gamma_x + \gamma') \cos\theta \\ Z_y \gamma_y (-\sin\theta) \end{pmatrix}, \quad \text{(S4)}$$

where we considered that (1) all emitted fields are equal to the integral over the charge current $I_{c,i}$ in the given direction multiplied by impedance $Z_i$, $i = x, y$, and the elementary charge $e$. Then, (2) $I_{c,i}$ is proportional to the ISHE conversion angle $\gamma_i$ and in the x-direction, also to the potential ISSE conversion angle $\gamma'$. Finally, (3) $I_{c,i}$ is always perpendicular to magnetization $\boldsymbol{M}$, oriented at an angle $\theta$ with respect to the crystal.

Assuming the anisotropy of the ISHE conversion angle, we can write $\gamma_x = \gamma_y + \delta\gamma$ and, similarly, anisotropic impedance can be written as $Z_x = Z_y + \delta Z$. Equation (S4) can be further expanded as:

$$\begin{pmatrix} (Z_y + \delta Z)(\gamma_y + \delta\gamma + \gamma') \cos\theta \\ Z_y \gamma_y (-\sin\theta) \end{pmatrix} = \begin{pmatrix} (Z_y\gamma_y + Z_y\delta\gamma + Z_y\gamma' + \delta Z\gamma_y + \delta Z\delta\gamma + \delta Z\gamma') \cos\theta \\ Z_y \gamma_y (-\sin\theta) \end{pmatrix} =$$

$$= Z_y\gamma_y \left[ \begin{pmatrix} \cos\theta \\ -\sin\theta \end{pmatrix} + \begin{pmatrix} \delta\gamma/\gamma_y + \gamma'/\gamma_y + \delta Z/Z_y + \Delta \\ 0 \end{pmatrix} \cos\theta \right], \quad \text{(S5)}$$

where $\Delta = (\delta Z/Z_y)(\delta\gamma/\gamma_y) + (\delta Z/Z_y)(\gamma'/\gamma_y)$.